\documentstyle[amssymb,aps,multicol,prl,epsf]{revtex}

\newcommand \beq  {\begin{equation}}
\newcommand \eeq  {\end{equation}}
\newcommand \bea {\begin{eqnarray} }
\newcommand \eea {\end{eqnarray}}

\begin{document}
\title{Dissipative dynamics of an
extended magnetic nanostructure : Spin necklace in a metallic environment}
\author{N. Shah$^{1}$ and A. J. Millis$^{2}$}
\address{$^{1}$Center for Materials Theory, Department of Physics \& Astronomy, Rutgers University,\\
136 Frelinghuysen Road, Piscataway, NJ 08854.\\
$^{2}$ Department of Physics, Columbia University,
538 W 120th St, NY, NY 10027.}
\maketitle
\begin{abstract}
We study theoretically the dynamics of an ``xxz'' spin necklace coupled to a conduction electron sea, a model system for a nanostructure in a dissipative environment. We extract the long-time behavior via a mapping to a multichannel Coulomb gas problem followed by a scaling analysis. The strong quantum fluctuations of the necklace cause a nontrivial dependence of couplings on system size which we extract via an analysis involving the ``boundary condition changing
operator'', and confirm via a detailed numerical evaluation of one case.
\end{abstract}
\begin{multicols}{2}

The dissipative dynamics of a magnetic nanostructure is important from both
technological and fundamental points of view. For example, magnetic
recording involves the polarization of small domains, whose stability over
long times is crucial. Tunnelling of isolated magnetic particles has
been extensively studied\cite{Chudnovsky}\cite{Gunther}. We shall be
interested here in the new physics brought by coupling to a conduction
electron bath. There are two effects. Exchange of spin between the
conduction electron system and the nanostructure changes the magnetic state
of the nanostructure, leading effectively to a tunneling process. On the
other hand, dissipation from the particle-hole excitations of the metal
suppresses tunnelling\cite{Leggett87}. A finite cluster of spins coupled to
a conduction electron bath is thus a model system for examining general
issues of quantum coherence and dissipation in a system with many degrees of
freedom. Also, recent experimental advances in atom manipulation\cite
{Eigler93}\cite{Eigler00} mean that clusters of atoms (eg. quantum corrals)
can now be placed in controllable arrays on suitably chosen solid surfaces,
so it seems likely that in the near future controlled magnetic nano-arrays
may be constructed and studied.

In the case of a single spin in a metal, the interplay between tunnelling
and dissipation gives rise to the logarithmic renormalizations
characteristic of the one-impurity Kondo effect\cite{Anderson69}. While the
two\cite{Jones87}\cite{Castro00} and three\cite{Paul96} impurity Kondo
problems have been studied, there is less known about the behavior of
systems containing larger numbers of spins. The subject has been recently
studied in the context of the ``Griffiths phase'' scenario for the
non-Fermi-liquid physics apparently observed in certain `heavy fermion'
metals\cite{Stewart01}. To study this physics, Castro-Neto and Jones\cite
{Castro00} considered a cluster of $N$ magnetic impurities, coupled strongly
to each other and weakly to a conduction electron bath. They presented a
series of qualitative arguments and approximate mappings from which they
concluded that the system could undergo a more or less conventional Kondo
screening (which they termed a ``cluster Kondo effect''), at an $N-$%
dependent temperature which they estimated. The qualitative nature of the
treatment suggests however that further analysis would be useful. One of us,
with Morr and Schmalian\cite{Millis01}, treated a similar problem of $N$
coupled moments from a functional integral point of view. In the Ising case,
orthogonality effects were found to suppress tunnelling in all but the
smallest nanostructures, while no definitive results were obtained in the
continuous symmetry case.

In this paper we attack the problem from a different point of view. We
identify a specific nanostructure, the finite-size spin necklace depicted in
Fig.1. There are three motivations for this choice of cluster. First, it is
(as can well be seen) amenable to a detailed analysis. Second, it exhibits
large (but tractable) quantum fluctuations whereas in other works\cite
{Castro00}\cite{Millis01} the quantum fluctuations within the cluster are
not taken into account. Third, we can treat the continuous symmetry case as
well.

\vspace{0.25cm} 
\centerline{\epsfxsize=2.2truein \epsfbox{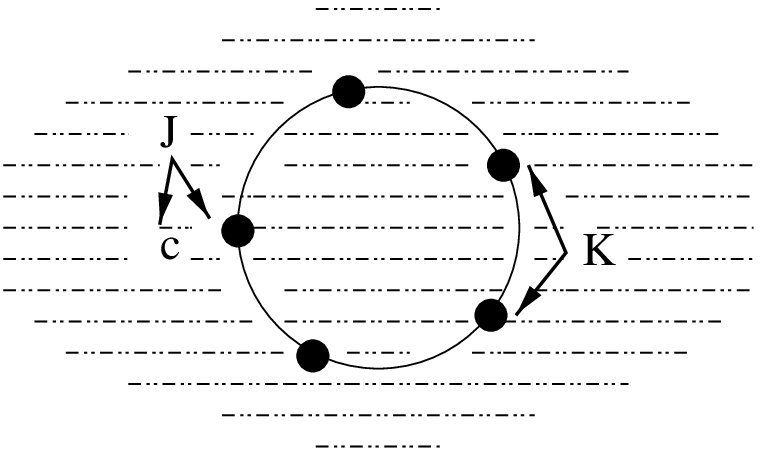}}
{\footnotesize \textbf{Fig.1} Sketch of `spin necklace' in conduction electron bath for $N=5$. Heavy dots represent local moments, which are coupled via an exchange coupling K. Shaded area represents conduction electron (c)
sea, Kondo-coupled to local moments via exchange J.}                    
\vspace{0.25cm}

The Hamiltonian describing the physics of interest is 
\begin{eqnarray}
H &=&H_{necklace}+H_{cond}+H_{coupling}^{z}+H_{coupling}^{\perp } \\
H_{necklace} &=&\sum_{j=0}^{N-1}(K_{\bot
}(S_{j}^{x}S_{j+1}^{x}+S_{j}^{y}S_{j+1}^{y})+K_{z}S_{j}^{z}S_{j+1}^{z}) \\
H_{cond} &=&\sum_{k}\epsilon _{k}c_{k\sigma }^{\dagger }c_{k\sigma } \\
H_{coupling}^{z} &=&J_{z}\sum_{j=0}^{N-1}S_{j}^{z}(c_{{\bf R}_{j}\uparrow \
}^{\dagger }c_{{\bf R}_{j}\uparrow \ }-c_{{\bf R}_{j}\downarrow \ }^{\dagger
}c_{{\bf R}_{j}\downarrow \ }) \\
H_{coupling}^{\perp } &=&J_{\bot }\sum\limits_{j=0}^{N-1}(S_{j}^{+}c_{{\bf R}%
_{j}\downarrow \ }^{\dagger }c_{{\bf R}_{j}\uparrow }+S_{j}^{-}c_{{\bf R}%
_{j}\uparrow }^{\dagger }c_{{\bf R}_{j}\downarrow })
\end{eqnarray}
with ${\bf R}_{j}$ the position of spin $j$. Here each spin value is $%
S_{j}=1/2$ and{\bf \ }${\bf S}_{N}={\bf S}_{0}$. We restrict to the case of $%
K_{z}>-\left| K_{\bot }\right| $, so the ground state of $H_{necklace}$ is
not ferromagnetic. We further specialize to $N$ odd so the ground state is
characterized by a spin quantum number $S^{z}=\pm 1/2$. For $K_{\bot }<0$
the total momentum of each of the $S^{z}=\pm 1/2$ states is zero and the
ground state is two-fold degenerate. For $K_{_{\bot }}>0$ the ground state
also has non-zero total momentum $P=\kappa \left| P\right| ,$ $\kappa =\pm 1$
implying four-fold degeneracy. In this paper we focus on the limit in which
the Kondo coupling, $J$, of the cluster to the conduction electrons is weak
enough and the temperatures of interest low enough, that it is sufficient to
consider only those spin flip processes that couple the degenerate ground
states of the cluster. We expect that stronger couplings, implying mixing of
higher energy states of the cluster, give rise to qualitatively similar
physics. This case will be considered in a future paper\cite{Shah03}.

If $H_{coupling}^{\perp }=0$, then the $z-$component of the cluster spin is
a good quantum number. The cluster remains in a definite spin state, the
conduction electrons just feel a potential due to the longitudinal coupling $%
J_{z}$, and the Hamiltonian $H_{0}=H_{necklace}+H_{cond}+H_{coupling}^{z}$
is easily diagonalized. For non-zero $J_{\bot }\ $ the problem is no longer
trivially solvable. To study it, we use a perturbative expansion in $J_{\bot
}$ to establish a mapping onto a Coulomb gas, following the approach
introduced by Anderson and Yuval\cite{Anderson69}. Let $\left| \Psi
_{\Uparrow }\right\rangle $ denote the ground state of $H_{0}$ when the
cluster is in the $S^{z}=+1/2$ state, $\left| \Uparrow \right\rangle $ and
consider the ground-state-to-ground-state amplitude (alternatively the
partition function in the imaginary time formulation) given in the
interaction representation by 
\begin{equation}
Z(t)=\left\langle \Psi _{\Uparrow }\left| e^{iH_{0}t}T\exp
[i\int\nolimits_{0}^{t}H^{\prime }(t^{\prime })dt^{\prime }]\right| \Psi
_{\Uparrow }\right\rangle
\end{equation}
where $H^{\prime }(t)=e^{-iH_{0}t}H_{coupling}^{\perp }e^{iH_{0}t}$ and$\ T$
is the ``time ordering operator''. Restricting to the ground state subspace
of the cluster, and expanding to ${\cal O}(H^{\prime 2})$, 
\begin{eqnarray}
Z(t) &=&J_{\bot
}^{2}\int\nolimits_{0}^{t}dt_{1}\int\nolimits_{0}^{t_{1}}dt_{2}\sum_{{\bf R}%
_{i}}\sum_{{\bf R}_{j}}\left\langle \Uparrow \right| S_{i}^{+}\left|
\Downarrow \right\rangle \left\langle \Downarrow \right| S_{j}^{-}\left|
\Uparrow \right\rangle  \nonumber \\
&&\times F_{\uparrow }({\bf R}_{ij},t_{12})F_{\downarrow }({\bf R}%
_{ij},t_{12})  \label{Z2}
\end{eqnarray}
where ${\bf R}_{ij}\equiv {\bf R}_{i}-{\bf R}_{j}$, $t_{12}\equiv
t_{1}-t_{2} $ and 
\begin{eqnarray}
F_{_{\uparrow }}({\bf R}_{ij},t_{12}) &=&\left\langle \Psi _{\Uparrow
}\left| Tc_{\uparrow {\bf R}_{i}}(t_{1})c_{\uparrow {\bf R}_{j}}^{\dagger
}(t_{2}){\cal S}(t_{12})\right| \Psi _{\Uparrow }\right\rangle \\
{\cal S}(t_{12}) &=&T\exp [-i\int\nolimits_{t_{2}}^{t_{1}}d\tau V(\tau )] \\
V(\tau ) &=&-2J_{z}\sum_{{\bf R}_{j}}\left\langle \Uparrow \right|
S_{j}^{z}\left| \Uparrow \right\rangle c_{{\bf R}_{j}}^{\dagger }(\tau )c_{%
{\bf R}_{j}}(\tau ).
\end{eqnarray}
The expressions above are similar to the ones introduced in the context of
the X-ray\cite{Nozieres69}\cite{Gogolin} and the Kondo problem\cite
{Anderson69}, but with new features arising from the spatially extended
nature of the magnetic object. The most important of these new features are
the space dependence of $F$, arising from processes in which an electron is
created at one site and destroyed at another, and a nontrivial dependence of
matrix elements such as $\left\langle \Uparrow \right| S_{i}^{+}\left|
\Downarrow \right\rangle $ on necklace size and symmetry. We discuss these
in turn.

Following Ref.\cite{Nozieres69} we express $F$ as the product of open line
part $g=F/\left\langle {\cal S}\right\rangle $ and a closed loop part $%
\left\langle {\cal S}\right\rangle $ $\equiv e^{C}$ which may be computed by
an expansion in $J_{z}$. The crucial object in this expansion is the Green's
functions $G({\bf r},t)$ of $H_{cond}$. We require the long time limit ($%
t\gg R/v_{F}$ with $R$ the radius of the necklace) in which $G({\bf r}%
,t)\rightarrow \rho _{F}G({\bf r})/it$ : for a spherical Fermi surface $G(%
{\bf r})=2\sin (k_{F}r)/v_{F}r$. In order to obtain a closed form for $g$ we
exploit the specific geometry of the necklace. We parametrize the positions
of the spins by the azimuthal angle $\phi _{j}=2\pi j/N,$ $j\in [0,N-1]$.
Using $R_{kl}=2R\sin (\phi _{kl}/2)$, where $\phi _{kl}\equiv \phi _{k}-\phi
_{l}$, $G(R_{kl})\rightarrow G(\phi _{kl})$. We finally define the
one-dimensional Fourier transform 
\begin{eqnarray}
G_{m} &=&\sum_{k-l=0}^{N-1}G(\phi _{kl})e^{im\phi _{kl}} \\
G(\phi _{kl}) &=&\frac{1}{N}\sum_{m}G_{m}e^{-im\phi _{kl}}
\end{eqnarray}
where we choose $m=-\frac{N-1}{2},...0,...\frac{N-1}{2}$ to make the
symmetry of the problem most apparent: $G_{-m}=G_{m}$. Using the $m-$basis
and defining $V=-2J_{z}\left\langle \Uparrow \right| S_{i}^{z}\left|
\Uparrow \right\rangle $,
\begin{eqnarray}
g(\phi _{ij},t_{12}) &=&\sum_{m}\frac{\rho _{F}G_{m}}{iN(t_{12})}e^{-im\phi
_{ij}}\left[ \frac{\tau _{c}}{t_{12}}\right] ^{2V\rho _{F}G_{m}} \\
e^{C(t_{12})} &=&\left[ \frac{\tau _{c}}{t_{12}}\right] ^{V^{2}\rho
_{F}^{2}\sum_{m}G_{m}^{2}}
\end{eqnarray}
where  $\tau _{c}$ is the short-time cut-off separating two spin-flips and
we have neglected terms of ${\cal O}$($J_{z}^{3}$) in the exponent.

We now consider the matrix elements. We first note that if $K_{\bot }>0$
then the necklace ground states $\left| \Uparrow \right\rangle $, $\left|
\Uparrow \right\rangle $ have a nonvanishing momentum $\pm P_{\Uparrow
,\Downarrow }$ and the matrix element depends on the momentum transfer: $%
\left\langle \Uparrow \right| S_{j}^{+}\left| \Downarrow \right\rangle
=\left\langle \Uparrow \right| S_{0}^{+}\left| \Downarrow \right\rangle \exp
[iQj]$ where $Q=(P_{\Downarrow }-P_{\Uparrow })$. When $K_{\bot }>0$, the
absolute value of the matrix element decays faster with $N$ when $%
P_{\Uparrow }=P_{\Downarrow }$. For $K_{\bot }>0$ when $P_{\Uparrow
}=-P_{\Downarrow }$ and for $K_{\bot }<0$ we find, defining $\alpha \leq 1$, 
\begin{eqnarray}
\left| \left\langle \Uparrow \right| S_{j}^{+}\left| \Downarrow
\right\rangle \right| &=&\left| \left\langle \Uparrow \right|
S_{0}^{+}\left| \Downarrow \right\rangle \right| \sim \frac{1}{N^{\alpha }}
\\
\left| \left\langle \Uparrow \right| S_{j}^{z}\left| \Uparrow \right\rangle
\right| &=&\left| \left\langle \Uparrow \right| S_{0}^{z}\left| \Uparrow
\right\rangle \right| \sim \frac{1}{N}.
\end{eqnarray}

For $\left| K_{z}\right| \leq \left| K_{\bot }\right| $, in terms of the
Jordan-Wigner fermion representation\cite{Lieb61}, the states $\left|
\Uparrow \right\rangle $ and $\left| \Downarrow \right\rangle $ have
different boundary conditions, one periodic and the other antiperiodic.
Considering the low energy form of the Hamiltonian given by the Luttinger
(massless)\ model and adopting the ``boundary-condition changing operator''
formalism\cite{Schotte69}\cite{Affleck94} we find 
\begin{equation}
\begin{tabular}{ccccc}
$\alpha =\frac{1}{2}(1-\frac{1}{\pi }\cos ^{-1}\frac{K_{z}}{\left| K_{\bot
}\right| })$ &  &  & for & $\left| K_{z}\right| \leq \left| K_{\bot }\right| 
$%
\end{tabular}
\end{equation}
which we verify by a detailed numerical analysis (see Fig.2) of the exact
lattice solution for the $K_{z}=0$ case ($xx-$model\cite{Lieb61}) for which $%
\alpha =1/4$.

\vspace{0.25cm} 
\centerline{\epsfxsize=3truein \epsfbox{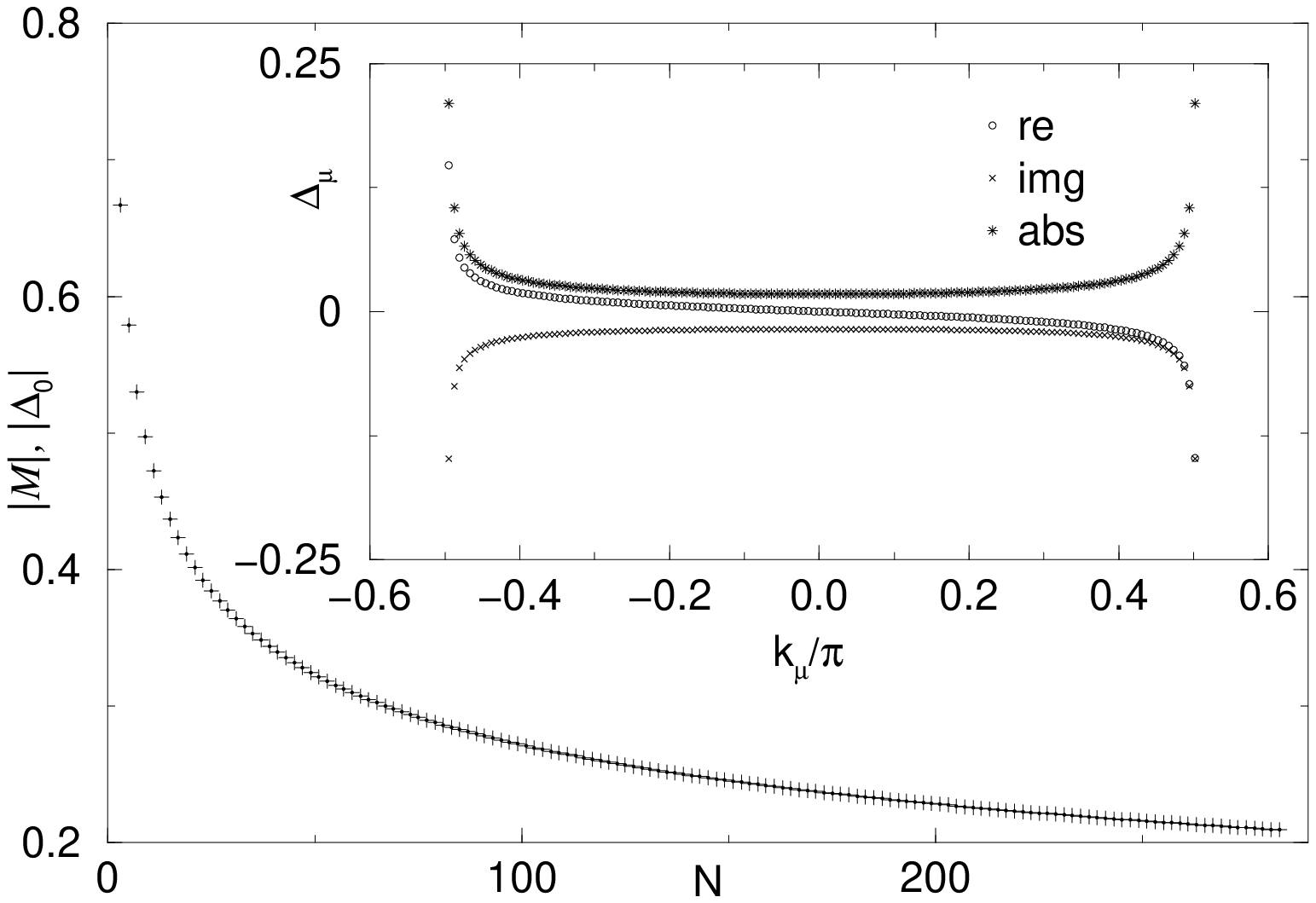}}
{\footnotesize \textbf{Fig.2} Plot of absolute value of $M=\left\langle \Uparrow \right| S_{j}^{+}\left| \Downarrow\right\rangle$ evaluated numerically for the $K_{z}=0$ case as a function of $N$. $\Delta _{\mu }$ is the matrix element when the Jordan-Wigner fermion is added (when spin is flipped) in the unoccupied momentum  $k_{\mu }$, each of which contributes to $M$. The inset plots its real, imaginary parts and the absolute value as a function of $k_{\mu }$ for $N=283$. The maximum contribution is near the Fermi point. We find that $\left|M\right|$ is equal to $\left|\Delta _{0}\right|$ (obtained when $k_{\mu }$ is closest ($\mu = 0$) to the Fermi point) and when plotted against $N$, the two plots overlap.}
\vspace{0.25cm}

For the Ising case ($K_{z}=\infty $) for $K_{z}>0,$ the energy is minimized
by having alternating up and down spins, but $N\;$being odd there is one
frustrated bond with $N\;$possible locations for each of $S^{z}=-\frac{1}{2}$
and $\frac{1}{2}$. A perturbative analysis around this $2N-$fold degenerate
point yields 
\begin{equation}
\begin{tabular}{ccccccc}
$\alpha =1$ &  &  &  &  & for & $K_{z}\gg $ $\left| K_{\bot }\right| .$%
\end{tabular}
\end{equation}
We expect $\alpha =1$ for all $K_{z}>\left| K_{\bot }\right| $ (Ising
regime).

The results for $\alpha $ indicate that it is much easier than one might
guess, to flip a large cluster between its two ground states. Quantum
fluctuations, which for example in the Ising regime delocalize the
frustrated bond, bring about dramatic enhancement in the tunnelling caused
by one spin flip by a conduction electron. By distributing $S^{z}=\pm 1/2$
over $N\;$sites, they also decrease the expectation value of $S_{j}^{z}$,
thereby reducing the potential felt by the conduction electrons at each spin
site.

We are now in a position to present our result for the
ground-state-to-ground-state amplitude for the spin necklace embedded in a
conduction bath, 
\begin{eqnarray}
Z(t) &=&\sum_{{\bf m}}u_{m}^{2}\int\nolimits_{0}^{t}\frac{dt_{1}}{\tau _{c}}%
\int\nolimits_{0}^{t_{1}}\frac{dt_{2}}{\tau _{c}}  \label{Z} \\
&&\times \exp \left\{ \left( 2-2\varepsilon _{m}\right) \ln \left[ \frac{%
\tau _{c}}{t^{\prime }-t}\right] \right\} 
\end{eqnarray}
where using $2\pi \widetilde{Q}/N=Q=(P_{\Downarrow }-P_{\Uparrow })$, 
\begin{eqnarray}
\varepsilon _{m} &=&\widetilde{J_{z}}[G_{m}+G_{m+\widetilde{Q}}]-\widetilde{%
J_{z}}^{2}\sum_{m^{\prime }}G_{m^{\prime }}^{2}  \label{em} \\
u_{m} &=&\widetilde{J_{\perp }}G_{m}G_{m+\widetilde{Q}}  \label{um}
\end{eqnarray}
with $\widetilde{J_{\perp }}=J_{\perp }\rho _{F}\left| \left\langle \Uparrow
\right| S_{0}^{+}\left| \Downarrow \right\rangle \right| \sim J_{\perp }\rho
_{F}/N^{\alpha }$ and $\widetilde{J_{z}}=2\rho _{F}J_{z}\left\langle
\Uparrow \right| S_{0}^{z}\left| \Uparrow \right\rangle \sim J_{z}\rho _{F}/N
$. Note that since $\sum_{m}G_{m}^{2}\sim N$, the two terms in $\varepsilon
_{m}$ are of the same order in $N$ and its leading behavior is given by the
term linear in $J_{z}\rho _{F}$ (assumed $\ll 1$). In the corresponding
expression obtained in Ref.\cite{Millis01} it was assumed that only one
channel was important and in the expression for $\varepsilon _{m}$, the term
linear in $J_{z}$ was overlooked. The above expression for $Z$, Eq.(\ref{Z}), is the second-order term in the expansion of a multicomponent Coulomb
gas model in which each spin flip event (Coulomb gas charge) is labelled by
the channel ($m-$value)\ of the electron-hole pair created. Higher order
terms in the expansion in $J_{\perp }$ may be treated similarly, but compact
expressions cannot be obtained because (as noted in a related context\cite
{Fabrizio95}) in the multi-channel case, the higher order terms do not
combine into the Cauchy determinant form\cite{Anderson69}. However we may
define a scaling procedure as usual\cite{Anderson70} by eliminating
close-pair spin flips of the same channel. To the leading order the
different channels are not coupled, and standard arguments show that near
the weak coupling fixed point, the partition function retains the form Eq.(%
\ref{Z}) but with parameters $\varepsilon _{m}(\tau _{c})$ and $u_{m}(\tau
_{c})$ which evolve according to 
\begin{eqnarray}
d\varepsilon _{m}/d(\ln \tau _{c}) &=&2u_{m}^{2}  \label{se1} \\
du_{m}/d(\ln \tau _{c}) &=&u_{m}\varepsilon _{m}/2  \label{se2}
\end{eqnarray}
with initial conditions given by Eq.(\ref{em}) and (\ref{um}). The flow
implied by Eq.(\ref{se1}) and (\ref{se2}) is shown in Fig.3. 

\vspace{0.25cm} 
\centerline{\epsfxsize=3truein \epsfbox{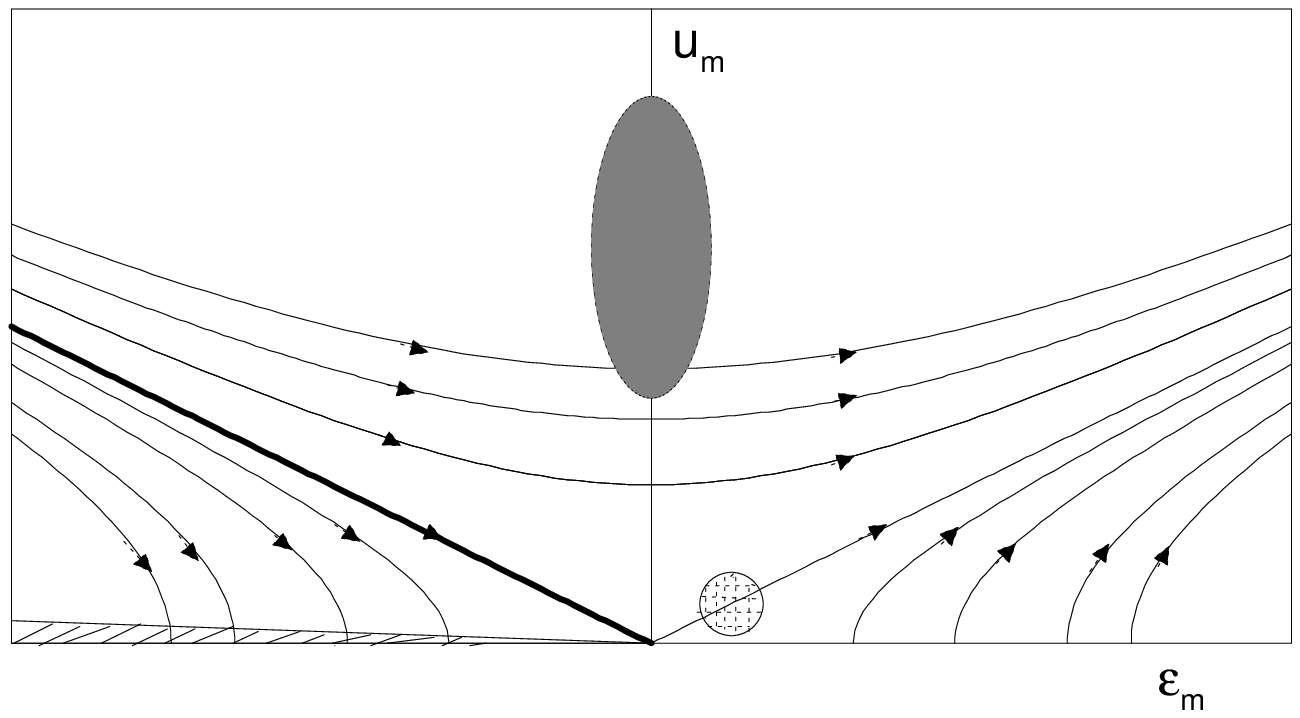}}
{\footnotesize \textbf{Fig.3} Plot showing the flow for each channel, near the weak coupling fixed point. The heavy solid line is the separatrix  $\varepsilon _{m}=-u_{m}/2$. The shaded oval region shows the generic initial conditions for the $xx$ symmetry. The dotted circle shows the generic initial conditions for the Ising limit. The striped region near $u_m=0$ is the range of initial conditions for `imperfect' necklaces.}
\vspace{0.25cm}

For each
channel the flow is to the no-flip fixed line ($u_{m}=0$) if $\varepsilon
_{m}<-u_{m}/2$ and towards the strongly coupled (rapidly flipping) regime
(with $\varepsilon _{m}=u_{m}/2$) if $\varepsilon _{m}>-u_{m}/2$. Scaling
amplifies the differences between channels. For example, in the $K_{\bot }<0$
case of the $xx-$model, the initial condition is $u_{m}^{0}=J^{\perp }\rho
_{F}G_{m}/N^{1/4}$ and $\varepsilon _{m}^{0}=2J_{z}\rho _{F}G_{m}/N$. For
large $N$, $\varepsilon _{m}^{0}$ is thus parametrically smaller than $%
u_{m}^{0}$; all initial conditions are near the vertical axis and the
dominant channels are those with the largest $G_{m}$. Representative results
are shown in Fig.4. Note that for the ideal ring studied here, $G_{-m}=G_{m}$%
. We see that all channels with $m<Nk_{F}/2\pi $ couple with roughly equal
strength, but generically there is one largest pair, so after a transient
many channel regime, the asymptotic behavior is of the two-channel Kondo
problem. Exceptions can occur, and in particular for $k_{F}\lesssim 2\pi /N$%
, the $m=0$ channel dominates. The same generic behavior occurs in other
cases we have studied, but we note that in the Ising regime the $J_{z}$ and $%
J_{\bot }$ couplings are both $\sim 1/N$, so the initial conditions are
shifted as indicated in Fig.3.

\vspace{0.25cm} 
\centerline{\epsfxsize=3truein \epsfbox{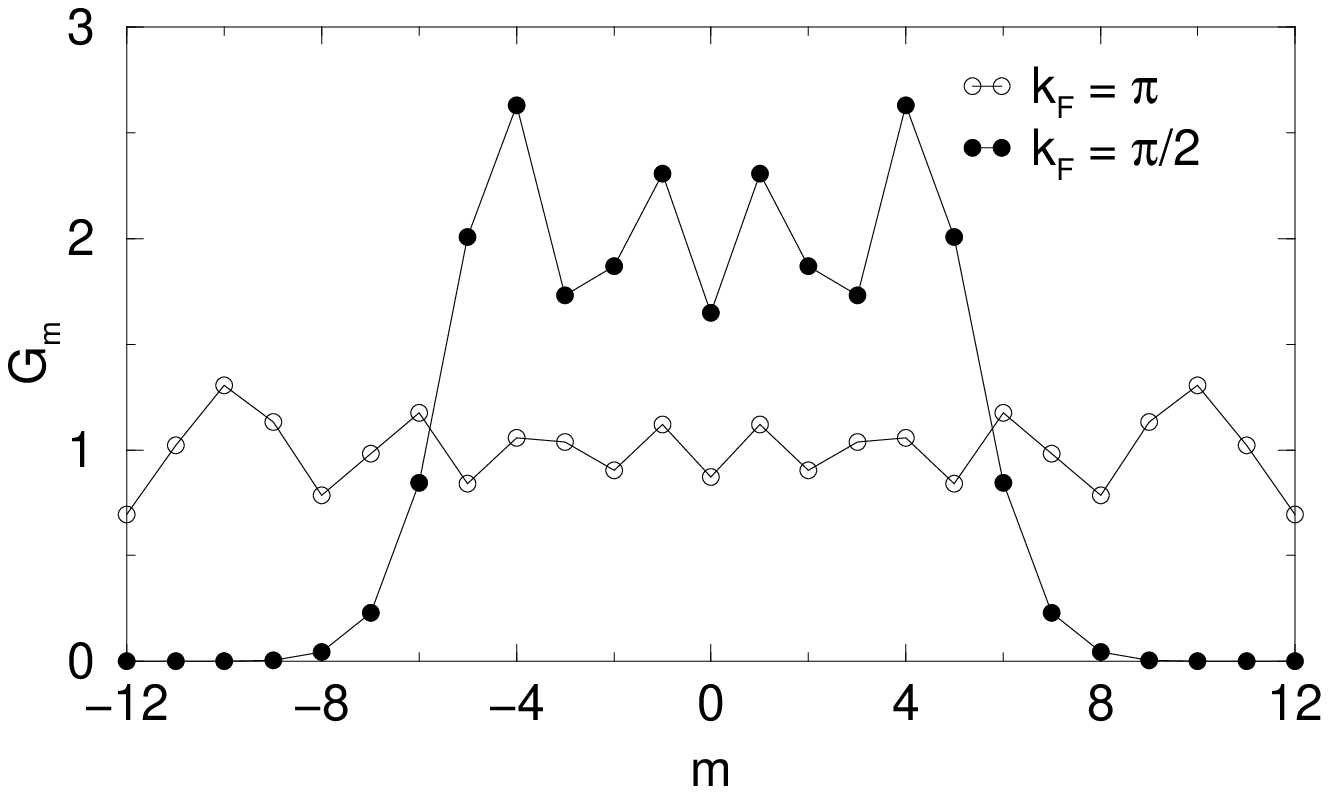}}
{\footnotesize \textbf{Fig.4} Plot showing  $G_m$ in the $K_{\bot }<0$ case of the $xx-$model for two values of $k_{F}$. For  $k_{F}=\pi$, the pair $m = -10$ and $10$  and for $k_{F}=\pi/2$, the pair $m = -4$ and $4$ is the highest, constituting the dominant pair which governs the low energy behavior.}
\vspace{0.25cm}

To summarize: we have shown that the generic long-time dynamics of a spin
necklace coupled to a conduction electron bath is of the two-channel Kondo
class. We emphasize that generic deviations from perfect translation
invariance break the two-channel symmetry. A key feature of the model is the
large quantum fluctuations characteristic of the isolated necklace: these
ensure that the dominant spin flip amplitude between the ground states of
the spin necklace decays only as a power of the system size ($N^{-\alpha }$
with $0<\alpha \leq 1$) and that the static field exerted by the necklace on
the conduction bath is small, of order $1/N,$ so ``orthogonality'' effects
do not dominate. To explore the changes occurring as quantum fluctuations
are reduced, we consider a necklace in the extreme Ising limit ($K_{z}\gg $ $%
\left| K_{\bot }\right| $) and with one weak link ($K_{z}^{01}\rightarrow
K_{z}(1-\delta )$) which acts to localize the frustrated bond. A detailed
discussion and extension to other cases will be published elsewhere\cite
{Shah03}. Here we note that although the residual reflection symmetry allows two
channel behavior, the spin flip amplitude changes from $\sim 1/N$ to $\sim
\exp [-N\ln (K_{\bot }/\delta K_{z})]$ (moving the initial condition much
closer to the $u_{m}=0$ axis of Fig.3). More importantly, because the
frustrated bond is localized, the quadratic, ``orthogonality'' term in $%
\varepsilon _{m}$ (Eq.(\ref{em})) becomes $\sim J_{z}^{2}N$, moving the
initial condition substantially to the left (independent of the sign of $%
J_{z}$); and, for large enough $N$, into the stable (no asymptotic flip)
region.

Extensions of the present work include a treatment of systems with open
boundary conditions, and of other nanostructures. We expect generically that
internal quantum fluctuations of a nanoscale object (neglected in many
treatments, Ref.\cite{Castro00} and \cite{Millis01} for example) will have a
dramatic effect on the long time dynamics.

Acknowledgments: We thank L. B. Ioffe for helpful discussions and
NSF-DMR-00081075 for support.

\end{multicols}
\end{document}